\documentclass[intlimits,twoside,a4paper]{article}

\usepackage{amsmath,amssymb}
\usepackage{graphicx}
\usepackage{wrapfig}
\usepackage{siunitx,color}
\usepackage{cite}

\usepackage[T2A]{fontenc}
\usepackage[cp1251]{inputenc}

\usepackage[eqsecnum]{cmpj2}

\newcommand{\eqb}{\begin{equation}\label{eq:\arabic{section}-\arabic{equation}}}
\newcommand{\eqe}{\end{equation}}

\issue{2013}{16}{1}{13401}

\doinumber{10.5488/CMP.16.13401}

\title[Conductivity and permittivity of dispersed systems]%
{Conductivity and permittivity of dispersed systems with
penetrable particle-host interphase}

\author{M.Ya.~Sushko, A.K.~Semenov}

\address{Mechnikov National University, Department of Theoretical
Physics, 2 Dvoryanska St., 65026 Odesa, Ukraine}%, E-mail: {mrs@onu.edu.ua} }

\authorcopyright{M.Ya.~Sushko, A.K.~Semenov, 2013}

\date{Received July 3, 2012, in final form November 28, 2012}

\begin{document}

\maketitle

\begin{abstract}
A model for the study of the effective quasistatic conductivity and
permittivity of dispersed systems with particle-host interphase,
within which many-particle polarization and correlation
contributions are effectively incorporated, is presented. The
structure of the system’s components, including the interphase, is
taken into account through modelling their low-frequency complex
permittivity profiles. The model describes, among other things,  a
percolation-type behavior of the effective conductivity,
accompanied by a considerable increase in the real part of the
effective complex permittivity. The percolation threshold location
is determined mainly by the thickness of the interphase. The
“double” percolation effect is predicted. The results are
contrasted with experiment.

\pacs 42.25.Dd, 64.60.Ak, 77.22.Ch, 82.70.-y, 83.80.Hj
\end{abstract}

\section{Introduction}

The studies of disperse systems, nanofluids, and systems of
nanoparticles are nontrivial, but important for many fields in
science and industry. Much attention has been recently focused on
the question how the properties of the particle-host interphase
affect the dielectric characteristics of the entire system~\cite{bib:Cai99,bib:Yan10,bib:Ran06,bib:Liu11}. This is an
essentially many-particle problem, playing a crucial role, say,
in percolation studies~\cite{bib:Kir73,bib:Cle90,bib:Sta03}. The
latter involve (a) various  numerical (mainly Monte-Carlo) methods
\cite{bib:Pik74,bib:Pik74a}, (b) graph~\cite{bib:Cal00} and
renormalization group~\cite{bib:Sta03} theories, (c) different
improvements of  the Maxwell-Garnett~\cite{bib:Max04} and
Bruggeman~\cite{bib:Bru35} approaches.

It should be noted that the practical realization of methods (a)
is labor-consuming and imposes high  computer specifications.
Methods (b) provide certain scaling relations, but ignore the
physical nature of dielectric response of an inhomogeneous system.
Methods (c) seem more adequate physically, but are actually
one-particle approaches. They can demonstrate a percolation-type
behavior of conductivity~\cite{bib:Sna07}, but ignore  the fractal
nature of percolation cluster~\cite{bib:Kig01}. Numerous attempts
at improving these approaches are widespread in modern
technical literature (see~\cite{bib:McL86}), but are, for most
part, poorly justified physically.

In the present work, the role of the interphase in the formation
of a percolation-type behavior of conductivity is analyzed within
an electrodynamic model based upon the method of compact groups of
inhomogeneities~\cite{bib:Su07,bib:Su09,bib:Su09a}. The ``compact
group'' means any macroscopic region, in which all distances
between the inhomogeneities are much shorter than the wavelength
of probing radiation in the medium. The method allows one to
effectively sum up and average the electric field and displacement
vectors in finely-dispersed systems of particles with complex
internal structure and arbitrary permittivities. In doing so,
unnecessary detailed elaboration of the processes of interparticle
polarization and correlations in the system is avoided.

The main idea of the method is that in the long-wave limit, a
macroscopically homogeneous and isotropic particulate system is
electrodynamically equivalent to an aggregate of compact groups of
particles. The compact groups include macroscopic numbers of
particles and  can be viewed as one-point inhomogeneities, with
negligibly small particle number fluctuations inside. They are in
fact self-similar. Note that this property is characteristic of  a
percolation cluster.

\begin{wrapfigure}{o}{0.48\textwidth}
  \centerline{\includegraphics[width=0.48\textwidth]{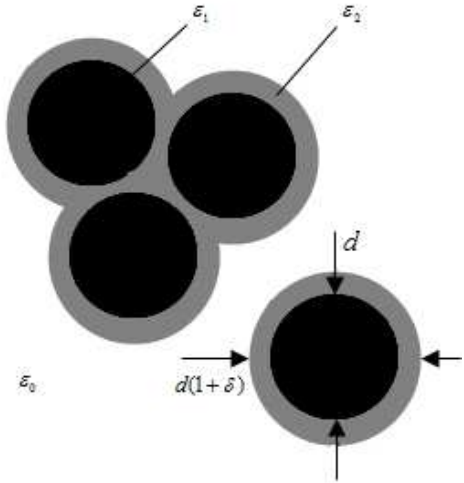}}
  \caption{The model  under consideration.  } \label{fig:1}
\end{wrapfigure}
The model of the disperse system under consideration  is depicted
in figure \ref{fig:1}. Each particle consists of  a spherical hard
core (black), with diameter $d$ and permittivity $\varepsilon_1$.
The core is surrounded by a penetrable concentric shell (gray),
with outer diameter $d(1+\delta)$ ($\delta$ being the relative
thickness of the shell) and permittivity $\varepsilon_2$. The
particles are embedded into a host medium (white), with
permittivity $\varepsilon_0$.  All the permittivities are in
general complex. The volume concentration of the hard-cores is
$c$. The shells can freely overlap; together, they form the
interphase between the hard cores and the host. The volume
concentration $\phi_\mathrm{int}$ of the interphase for particular values $c$
and $\delta$ can be calculated by statistical methods. We take it
from Monte Carlo simulations~\cite{bib:Rik85,bib:Rot03}.

We expect that, in the main features,  our results should remain
valid for macroscopically homogeneous and isotropic systems with
non-spherical core-shell particles: the sphericity comes into play
at the points where the explicit form of the function
$\phi_\mathrm{int}=\phi_\mathrm{int}(c,\delta)$ is needed. By contrast, spherical symmetry
of the particles is crucial for most approaches to the problem
which often represent different versions of one- and two-particle
approximations and use it either implicitly or through the
requirement that the shells should be hard. On the other hand, the
systems of  anisotropic particles are usually treated as if the
latter were solitary. Some examples and further literature on
these topics can be found in~\cite{bib:Su09a,bib:Tor84,bib:Nan93,bib:Liu11}.

\bigskip

\section{General equation for the low-frequency complex permittivity}
The structure of the system’s components, including the
interphase, is taken into account through modelling their complex
permittivity profiles and the choice of the way of homogenization.

We are interested in the low-frequency value $\varepsilon =
\varepsilon' - {\rm i}{4\pi \sigma} /{\omega } $ of the effective
complex permittivity and determine it as the proportionality
coefficient in the relation
\begin{equation} \label{eq0}
\langle \bf{D} (\bf{r})\rangle = \langle \epsilon (\bf{r}) \bf{E}
(\bf{r}) \rangle = \varepsilon \langle \bf{E} (\bf{r}) \rangle,
\end{equation}
where $\bf{D} (\bf{r})$, $\bf{E} (\bf{r})$, and $\epsilon
(\bf{r})$ are the local values of, respectively, the electric
displacement, electric field, and permittivity in the system, and
the angular brackets stand for the statistical averaging. In the
present work, we suggest that  this averaging can be reduced to
the following two-step procedure.
\begin{enumerate}
\item Finding the equilibrium distribution of the particles with
penetrable outer shells. This distribution is characterized by the
hard-cores' volume concentration $c$ and the particles' effective
volume concentration $\phi = \phi(c,\delta)$ (the latter is the
sum of $c$ and the volume concentration of the regions occupied by
the shells). Electrodynamically, such a system can be viewed as an
aggregate of non-overlapping (white, black, and gray in figure~\ref{fig:1}) regions with permittivities $\varepsilon_i =
\varepsilon'_i - {\rm i}{4\pi \sigma_i}/ {\omega } $, $i=0,1,2$,
and volume concentrations $1-\phi$, $c$, and $\phi -c$,
respectively.
\item
Electrodynamic homogenization, by direct integration with
respect to its volume~\cite{bib:Lan84}, of this aggregate within
the compact group approach. This suggests that: (a) with respect
to a probing field of long wavelength $\lambda$, the aggregate
itself is viewed  as a set of regions (termed as compact groups)
with typical linear sizes much smaller than $\lambda$, but yet
remaining macroscopic; (b) the macroscopic dielectric properties
of the aggregate are equivalent to those of a macroscopically
homogeneous and isotropic system prepared by embedding the
aggregate's non-overlapping  regions into some fictitious medium,
of permittivity $\epsilon$. The permittivity distribution in this
fictitious system can be written as
\begin{equation}
\label{eq1} \epsilon  ({\rm {\bf r}}) = \epsilon  + \delta
\epsilon({\rm {\bf r}}),
\end{equation}
where, in our case,
\begin{equation} \label{eq37}  \delta \epsilon  ({\rm {\bf r}}) = (\varepsilon_{0}
 - \epsilon) \,\Pi_0 ({\rm {\bf r}}, \Omega _0)+ \sum\limits_{a =
 1}^{N} (\varepsilon _{1}  - \epsilon)\,\Pi_1 ({\rm {\bf r}},{\rm
 \Omega}_{a} ) + \sum\limits_{b = 1}^{N'} (\varepsilon _{2}  -
 \epsilon)\,\Pi_2 ({\rm {\bf r}},{\rm \Omega}_{b}' )
\end{equation}
is  the local permittivity deviation from $\epsilon$  due to the
presence of a compact group of the aggregate's non-overlapping
regions  at point ${\rm {\bf r}}$. Here,  $\Pi_0\left( {\bf r},
\Omega _0\right)$, $\Pi_1\left( {{\rm {\bf r}},\Omega _{a} }
\right)$, and $\Pi_2\left( {{\rm {\bf r}},\Omega _{b} '} \right)$
are the charac\-te\-ris\-tic functions of the regions $\Omega_0$,
$\Omega_a$, and $\Omega'_b$ occupied by, respectively, the real
host, the $a$th particle's hard core, and the $b$th connected
cluster of the overlapping outer shells:
\begin{equation}
\label{eq33} \Pi_i ({\rm {\bf r}},\Omega )=
\begin{cases} 1\,, & {\rm {\bf r}}\in {\rm \Omega}, \\
 0\,, & {\rm {\bf r}}\not \in {\rm \Omega}.
\end{cases}
\end{equation}
\end{enumerate}

With  the model permittivity distribution (\ref{eq1}) and
(\ref{eq37}) proposed, the averaged displacement and field in
formula (\ref{eq0}) are calculated by the general formulae
\begin{equation} \label{eq5} \langle{\rm {\bf  {E}}}\rangle =
{\left\{ {1 + {\sum\limits_{s = 1}^{\infty}  {\left( { -
{\frac{{1}}{{3\epsilon }} }} \right)^{s} \langle {{\mathop {\left[
{\delta \epsilon ({\rm {\bf r}})} \right]^{s}}}}} }}\rangle
\right\}}\,{\rm {\bf E}}_{0}\,,
\end{equation}
\begin{equation}\label{eq6}
\langle{\rm {\bf {D}}}\rangle = {\left\{ {\epsilon  + \epsilon
{\sum\limits_{s = 1}^{\infty}  {\left( { - {\frac{{1}}{{3\epsilon
}}}} \right)^{s}{\langle {\mathop {\left[ {\delta \epsilon ({\rm
{\bf r}})} \right]^{s}}}}\rangle} }  + {\sum\limits_{s =
0}^{\infty}  {\left( { - {\frac{{1}}{{3\epsilon }} }}
\right)^{s}{\langle {\mathop {\left[ {\delta \epsilon ({\rm {\bf
r}})} \right]^{s + 1}}}}\rangle} }}  \right\}}\,{\rm {\bf E}}_{0}\,.
\end{equation}

In view of the property (\ref{eq33}), the averages
$$\langle\left[ {\delta \varepsilon ({\rm {\bf r}})}
\right]^{n}\rangle =\frac{1}{V} \int\limits_V \left[ {\delta
\varepsilon ({\rm {\bf r}})} \right]^{n}{\rm d}{\rm {\bf r}}$$ are
easy to find:
\begin{equation}
\label{eq58} \langle \left[ {\delta \epsilon ({\rm {\bf r}})}
\right]^{n}\rangle = (1 - \phi )(\varepsilon _{0} - \epsilon )^{n}
+ c\left( {\varepsilon _{1} - \epsilon} \right)^{n} + \left( {\phi
- c} \right)(\varepsilon _{2} - \epsilon )^{n}.
\end{equation}
Substitution of (\ref{eq58}) into series (\ref{eq5}), (\ref{eq6})
and summation of those give
\begin{equation}
\label{eq61} (1 - \phi )\frac{\varepsilon _{0} - \epsilon
}{2\epsilon + \varepsilon _{0}}  + c\frac{\varepsilon _{1} -
\epsilon} {2\epsilon + \varepsilon _{1}} + (\phi -
c)\frac{\varepsilon _{2} - \epsilon} {2\epsilon + \varepsilon _{2}
} = \frac{\varepsilon  - \epsilon} {2\epsilon + \varepsilon }\,.
\end{equation}

The permittivity $\epsilon$ is the only unknown parameter left. In
what follows, we suggest that the dielectric properties of the
fictitious and effective systems are identical, that is,
$\epsilon=\varepsilon$. This approximation is equivalent to the
Bruggeman-type homogenization, when the host and dispersed
particles are treated symmetrically. This seems reasonable for the
system under study because the shape of the white and gray regions
is extremely complicated. The degree of this complexity increases
with $c$ even more.

The other extreme case, $\epsilon = \varepsilon_0$, corresponds to
the Maxwell-Garnett type of homogenization.

\section{Main features of the model}

In the quasistatic limit and at $\epsilon=\varepsilon $, the
general equation (\ref{eq61}) reduces to two real equations, a
cubic one for $\sigma$ and, once  $\sigma$ is found, a linear for
$\varepsilon'$:
\begin{equation} \label{cond1} (1 - \phi )\frac{\sigma _{0} -
\sigma }{2\sigma + \sigma_{0}}  + c\frac{\sigma_{1} - \sigma}
{2\sigma + \sigma_{1}} + (\phi - c)\frac{\sigma_{2} - \sigma}
{2\sigma + \sigma_{2} } = 0.
\end{equation}
\begin{eqnarray} \label{perm1} (1 - \phi ) \frac{\varepsilon'_{0} \sigma-
\varepsilon' \sigma_{0}}{(2\sigma + \sigma_{0})^2}+ c
\frac{\varepsilon'_{1} \sigma- \varepsilon'\sigma_{1} }{(2\sigma +
\sigma_{1})^2}  + (\phi - c) \frac{\varepsilon'_{2}\sigma -
\varepsilon' \sigma_{2}}{(2\sigma + \sigma_{2})^2} = 0.
\end{eqnarray}

A general analytical analysis of equation (\ref{cond1}) can be
carried out with the use of Cardano's formulas and is cumbersome \cite{bib:Kor68}. Nonetheless, the major features of the model can
be grasped with simple physical reasoning. For the sake of
convenience, we change in equations (\ref{cond1}) and
(\ref{perm1}) to dimensionless variables $x \equiv \sigma
/\sigma_1$, $y= \varepsilon'/\varepsilon_0'$ and $x_i \equiv
\sigma_i /\sigma_1$, $y_i= \varepsilon'_i /\varepsilon_0'$ ($i=
0,1,2$). Typically, $x_0 \ll 1$ and $y > 1$.

\subsection{Percolation-type behavior}
\begin{figure}[ht]
\centerline{\includegraphics[width=0.5\textwidth]{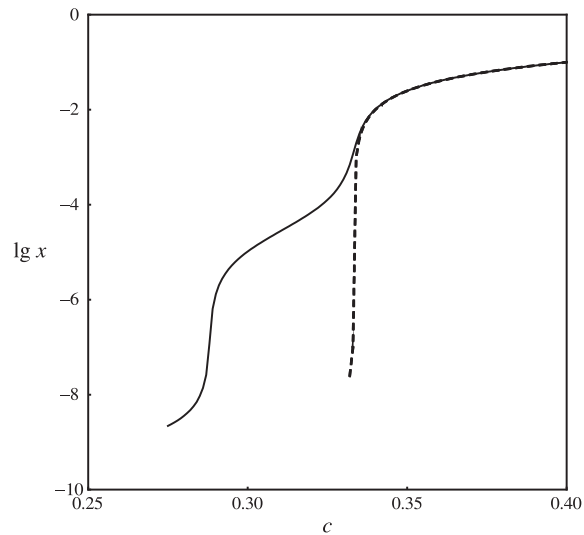}}
\caption{Percolation (dashed line, $\delta =0$) and ``double''
percolation (solid line, $\delta =0.05$); $x_0=1\cdot 10^{-10}$,
$x_2=5\cdot 10^{-5}$.} \label{fig:Perc}
\end{figure}

In the limit of a non-conducting matrix, $x_0 \to 0$, equation
(\ref{cond1}) has three solutions, $x =0$ and
\begin{equation} \label{thr0} x=\frac{3}{4}\left\{\left(c-\frac{1}{3}\right) +
 \left(\phi - \frac{1}{3}-c\right) x_2 \pm
\sqrt{\frac{4}{3}\left(\phi-\frac{1}{3}\right)x_2+
\left[\left(c-\frac{1}{3}\right) + \left(\phi - \frac{1}{3}
-c\right) x_2 \right]^2} \right\}.
\end{equation}
If $x_2>0$, a physically meaningful nontrivial solution (that with
the positive sign in front of the square root, see figure~\ref{fig:Perc}) appears only under the condition
\begin{equation} \label{thr1} \phi(c_{\rm c},\delta)=\frac{1}{3}
\end{equation} and is independent of $x_2$.
In the case $x_2 = 0$, the well-known value of $c_{\rm c}= 1/3$
for the two-component Bruggeman model occurs~\cite{bib:Sna07}.

The relation (\ref{thr1}) determines the threshold concentration
$c_{\rm c}$ for the effective conductivity. This value is affected
only by the fact of the existence  of the interphase ($\delta \neq
0$), rather than its dielectric and conductive properties.

\begin{figure}[ht]
\centerline{\includegraphics[width=0.47\textwidth]{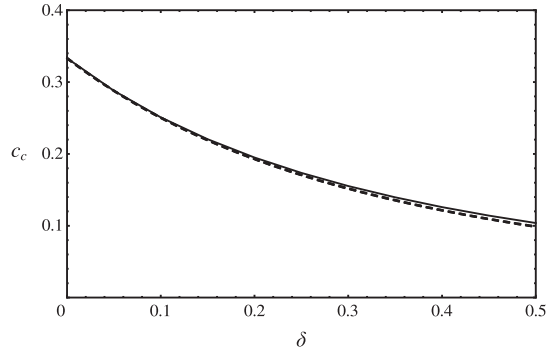}}
\caption{Threshold concentration as a function of the relative
thickness of the interphase (solid line). Dashed line: $c_{\rm c}
\simeq \frac{1}{3}(1+\delta)^{-3}$.} \label{fig:Thr}
\end{figure}

Our estimate of $c_{\rm c}$ as a function of the relative
thickness $\delta$ of the interphase layer  is shown in figure~\ref{fig:Thr}. The calculations were performed using the Monte
Carlo results~\cite{bib:Rot03} ($k = (1 + \delta )^{ - 3}$,
$\varphi = {{{c}}/{{k}}}$):
\begin{eqnarray}
 \phi &=& 1 - (1 - k\varphi )\exp {\left[ { - (1 - k)\varphi
 } \right]}\exp {\Bigg\{ { - {\frac{{k\varphi
 ^{2}}}{{2(1 - k\varphi
 )^{3}}}}{\left[ {(8 - 9k^{1 / 3} + k)} \right.}} } \nonumber\\
&&{} - (4 + 9k^{1 / 3} - 18k^{2 / 3} + 5k)k\varphi  + 2(1 -
 k)k^{2}{ {{\left. {\varphi^{2}} \right]}} \Bigg\}}.
 \label{eq8}
\end{eqnarray}
The analysis revealed that for realistic $c \lesssim 0.5$, the
relation $c_{\rm c} \simeq \frac{1}{3}(1+\delta)^{-3}$ can be
used.

In the immediate vicinity of $c_{\rm c}$ ($c \rightarrow c_{\rm
c}+0$) and for nonzero $\delta$, formula (\ref{thr0}) takes the
form
\begin{equation} \label{thr4} x\simeq\frac{3}{4} x_2\left[1+
\frac{\frac{1}{3}+c(1-x_2)}{\frac{1}{3}-c(1-x_2)}\right]\left(\phi
- \frac{1}{3}\right).
\end{equation}
Correspondingly, the effective conductivity  $\sigma \propto
(c-c_{\rm c})^t$ where the critical exponent $t\simeq 1.0$. The
effective permittivity $\varepsilon'$, as follows from equation
(\ref{perm1}), increases anomalously as $x_0 \to 0$. The latter
fact is in accord with predictions~\cite{bib:Efr76}.

\subsection{Effective critical exponent of conductivity}
In practice, both the threshold concentration $c_{\rm c}$ and the
critical exponent $t$ are determined by interpolating the
conductivity experimental data $\sigma = \sigma (c)$, obtained for
some finite interval $c \in [c_1, c_2]$ near $c_{\rm c}$ ($c_1 >
c_{\rm c}$), with the scaling law $\sigma =A (c-c_{\rm c})^t $,
$A$ and $t$ being independent of $c$. Then,
\begin{equation} \label{ind1}
t= {\log \frac{\sigma (c_2)}{\sigma (c_1)}}\bigg/{\log \frac{c_2
-c_{\rm c}}{c_1 -c_{\rm c}}}\,,
\end{equation} and this value is expected to be a $c$-independent constant.

\begin{figure}[!b]
\centerline{\includegraphics[width=0.47\textwidth]{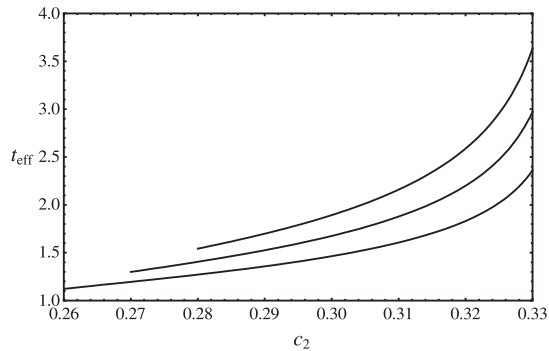}}
\caption{Effective critical exponent of conductivity as a function
of $c_2$ at fixed $c_1$, $\delta = 0.1$ ($c_{\rm c}\simeq 0.251$),
and $x_2=5\cdot 10^{-5}$, calculated with formulas (\ref{thr0})
and (\ref{ind1}). From bottom to top, $c_1 =0.26$, $0.27$, and
$0.28$.}  \label{fig:Ind}
\end{figure}
By contrast,  the asymptotic formula (\ref{thr4}) reveals that
even small variations of $c$ near $c_{\rm c}$ cause the expression
in the brackets, which is proportional to $A$, to change
considerably. This means that  a formal application  of the above
procedure and formula (\ref{ind1}) to a system with conductivity
(\ref{thr0}) will result in an effective exponent $t_{\rm eff} $
sensitive to the parameters $c_1$ and $c_2$ (figure~\ref{fig:Ind}). In particular, for a given $\delta \neq 0$,
$t_{\rm eff} $ increases as the interval $[c_1, c_2]$ (where
$c_2<1/3$) is: (a) shifted to higher values of $c$ (while its
width is fixed); (b) widened at fixed $c_1$. Also, the threshold
concentration found within this procedure is expected to exceed
$c_{\rm c}$.

Different 3D percolation models~\cite{bib:Kir73} and
renormalization group calculations~\cite{bib:Ber78,bib:Luc85} give
for $t$ estimates of $ \approx 1.3 \div 1.7$ and $\approx 1.9$,
$2.14$, respectively. Experimental values of $t$ are usually $1.5
\div 2.0$ and sometimes even twice as much~\cite{bib:Nan93}. As
evident from figure \ref{fig:Ind}, our theory is capable of
reproducing a variety of these values.

\subsection{Effect of the matrix's conductivity}
\begin{figure}[!t]
\begin{minipage}[h]{0.49\textwidth}
\centerline{\includegraphics[width=\textwidth]{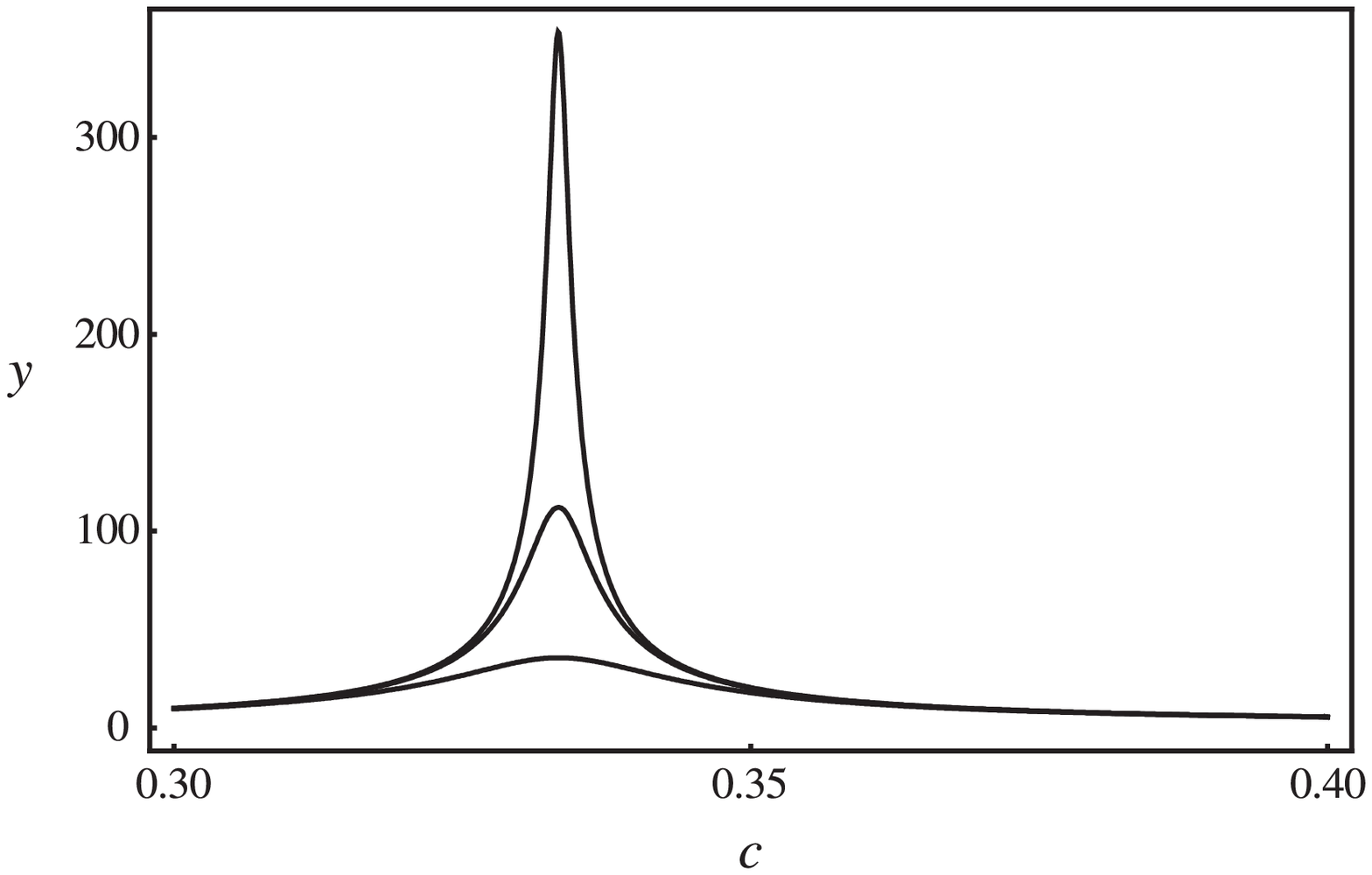}}
\caption{Effect of the matrix's conductivity on the effective
permittivity. From top to bottom, $x_0=1\cdot 10^{-6}$, $1\cdot
10^{-5}$, and $1\cdot 10^{-4}$. The other parameters: $y_1= 1.5$,
$y_2= 1$, $x_2= 0.05$, $\delta =0.05$.} \label{fig:Max}
\end{minipage}
\hfill
\begin{minipage}[h]{0.49\textwidth}
\centerline{\includegraphics[width=\textwidth]{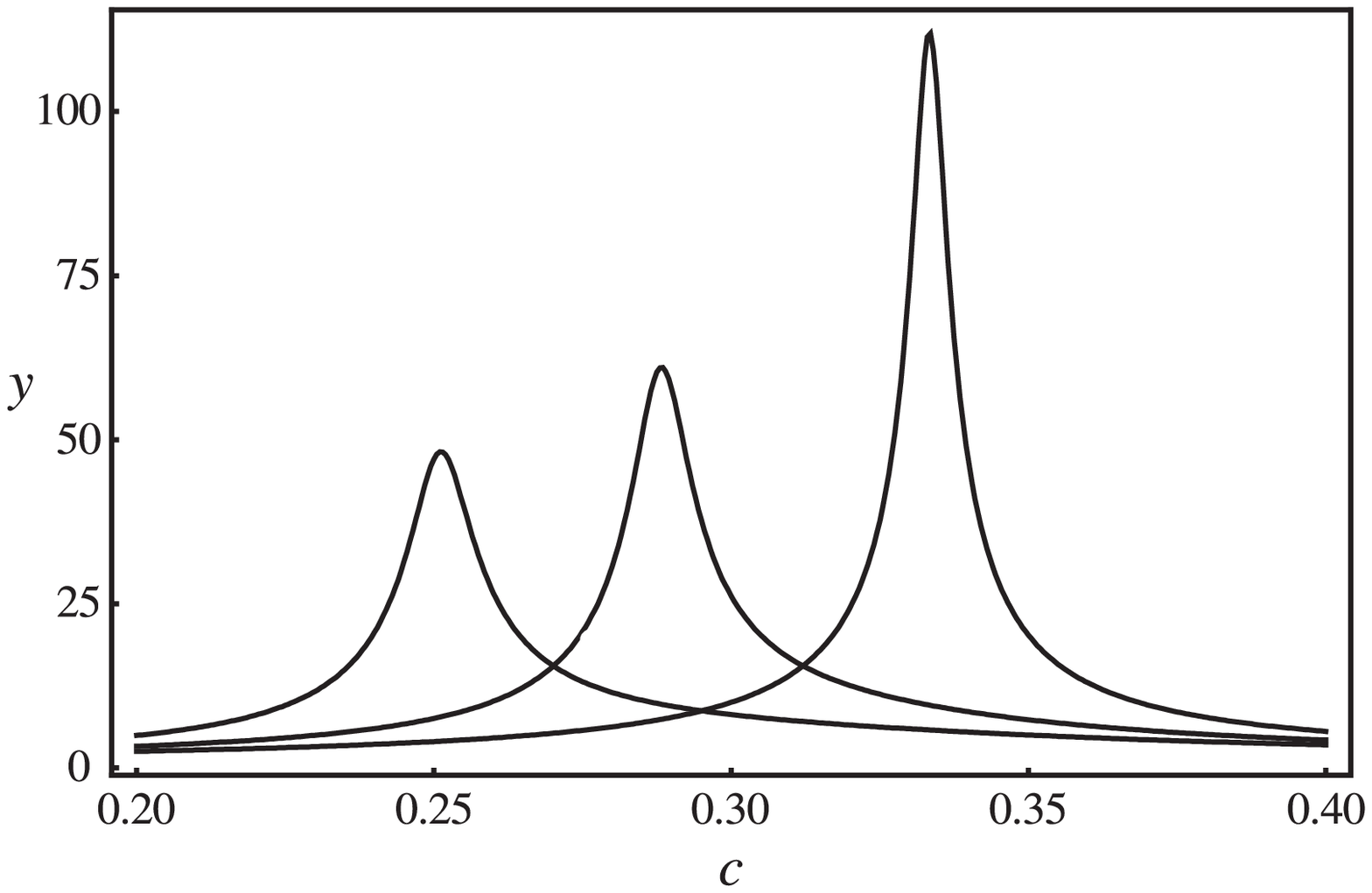}}
\caption{Effect of the interphase thickness on the effective
permittivity. From right to left, $\delta=0$, $0.05$, and $0.10$.
The other parameters: $y_1= 1.5$, $y_2= 1$, $x_0=1\cdot 10^{-5}$,
$x_2= 0.05$.} \label{fig:Loc}
\end{minipage}
\end{figure}

For real substances, $x_0 \neq 0$, though $x_0$ can be extremely
small. As a result, the percolation-type dependence of $x$ with
$c$  changes to a smooth one, with its slope  considerably
increasing  near $c_{\rm c}$. Simultaneously, the maximum value of
$y$ becomes bounded above and decreases as $x_0$ increases (figure~\ref{fig:Max}). The location of the maximum shifts to lower
concentrations as $\delta$ increases (figure~\ref{fig:Loc}).
Calculations show  that it remains practically independent of
$x_2$ and actually equal to $c_{\rm c}$.

\begin{figure}[!b]
\centerline{\includegraphics[width=0.48\textwidth]{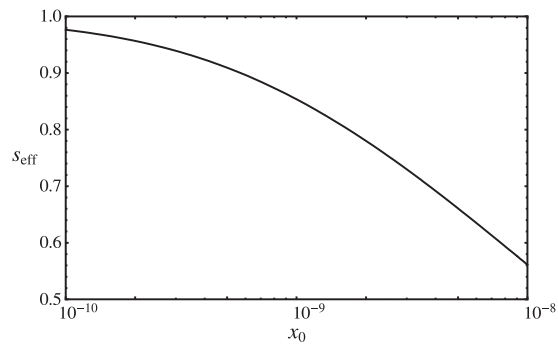}}
\caption{Effective critical exponent of conductivity below $c_{\rm
c}$ as a function of $x_0$ for $\delta = 0.1$ ($c_{\rm c}\simeq
0.251$) and $x_2=5\cdot 10^{-5}$, calculated with formulas and
(\ref{cond1}) and (\ref{ind2}) at $c_1 =0.24$ and
 $c_2=0.25$.} \label{fig:Ind1}
\end{figure}

Below the percolation threshold, the effective conductivity is
usually approximated by the scaling law $ \sigma =B (c_{\rm
c}-c)^{-s} $. Given experimental data for some interval $c \in
[c_1, c_2]$, $c_2<c_{\rm c}$, the effective values $s_{\rm eff}$
of  the critical exponent $s$ are found as
\begin{equation} \label{ind2}
s= - {\log \frac{\sigma (c_2)}{\sigma (c_1)}}\bigg/{\log
\frac{c_{\rm c}-c_2}{c_{\rm c}-c_1 }}\,.
\end{equation}

Our estimates of $s_{\rm eff}$ with formulas (\ref{cond1}) and
(\ref{ind2}) are shown in figure~\ref{fig:Ind1}. They correlate
well with typical theoretical~\cite{bib:Ber78,bib:Luc85} and
experimental~\cite{bib:Nan93} values of $0.75$ and $0.7 \div 1.0$,
respectively.

\subsection{Behavior of the permittivity}

According to equation (\ref{perm1}), the effective permittivity is
given by
\begin{equation} \label{perm2}
\varepsilon' =x \frac{(1-\phi)\, \varepsilon'_0 +
c\cfrac{(2x+x_0)^2}{(2x+1)^2}\,\varepsilon'_1 +(\phi
-c)\cfrac{(2x+x_0)^2}{(2x+x_2)^2}\,\varepsilon'_2}{(1-\phi) \, x_0
+ c\cfrac{(2x+x_0)^2}{(2x+1)^2} +(\phi
-c)\cfrac{(2x+x_0)^2}{(2x+x_2)^2}\, x_2}\,.
\end{equation}

For a badly-conducting matrix ($x_0 \rightarrow 0$) and under the
condition $x \ll 1$, three particular situations are of interest
to point out to.
\begin{enumerate}
\item The system is below the percolation threshold and the
conditions $x \ll \sqrt{x_0} $, $x \ll \sqrt{x_0 x_2}$, $x \ll {
x_2}$ (that is, $\sigma \ll \sqrt{\sigma_0 \sigma_1}$, $\sigma \ll
\sqrt{\sigma_0 \sigma_2}$, and $\sigma \ll {\sigma_2}$) are met.
Then, the dominant contributions to both the numerator and the
denominator are made by their first terms, and we expect that
$\varepsilon' \propto x \propto (c_{\rm c} -c )^{-s_{\rm eff}}$.
\item The system is above the threshold and $x \gg \sqrt{x_0}$, $x
\gg \sqrt{x_2}$, $x \gg {x_2}$ ($\sigma \gg \sqrt{\sigma_0
\sigma_1}$, $\sigma \gg  \sqrt{\sigma_1 \sigma_2}$, and $\sigma
\gg {\sigma_2}$). Now, of significance become the first and the
third terms in the numerator, the latter being almost independent
of $x$, and the second term in the denominator. Correspondingly,
the $\varepsilon'$ versus $c$ dependence is expected to be close
to $\varepsilon' \propto x^{-1} \propto (c - c_{\rm c} )^{-t_{\rm
eff}}$, with the proportionality constant slightly dependent on
$c$.

The exponents $s_{\rm eff}$ and $t_{\rm eff}$ in the two preceding
scaling-like laws are independent of the components'
permittivities $\varepsilon'_i$.
\item
The system is close to the percolation threshold, $ x \gg
\sqrt{x_0} $, and $ x \gg x_2 $ ($\sigma \gg \sqrt{\sigma_0
\sigma_1}$, $\sigma \gg \sigma_2$). Then, the numerator is almost
$x$-independent, whereas the denominator is mainly contributed to
by the second and the third terms. The dependence of
$\varepsilon'$ on $x$ takes the form $\varepsilon' \propto
ax/\left(1+bx^2\right)$, where the coefficients $a$ and $b$ are
easy to recover.
\end{enumerate}

\subsection{Applicability to real systems}
\begin{figure}[!t]
\centerline{\includegraphics[width=7cm]{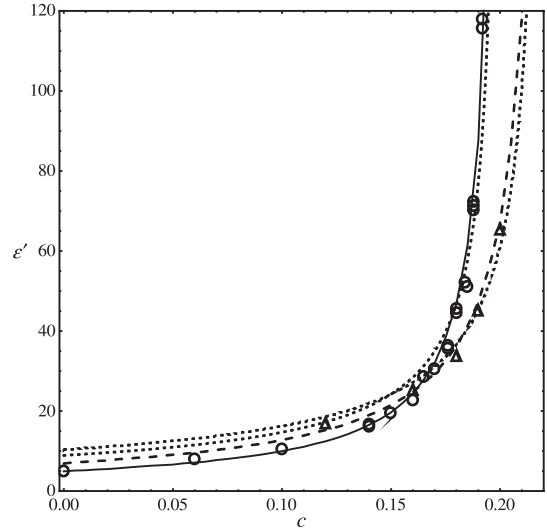}}
\caption{Effective permittivity data~\cite{bib:Gra81} for two
series of Ag--KCl composite samples (circles and triangles) below
the percolation threshold  and their fits with formula
(\ref{perm2}) at $\varepsilon_0' =5.0$, $\delta = 0.186$ (solid
line) and $\varepsilon_0' =7.0$, $\delta = 0.145$ (dashed line).
The dotted lines are the scaling-type fits (with $ c_{\rm c} =0.20
$, $s_{\rm eff} = 0.72 $ and $ c_{\rm c} =0.22 $, $s_{\rm eff} =
0.74$, respectively) proposed in~\cite{bib:Gra81} to the data for
$c>0.11$. } \label{fig:Permit3d}
\end{figure}

Figure~\ref{fig:Permit3d} shows the results of processing with
 formula (\ref{perm2}) the experimental data
\cite{bib:Gra81} for the effective permittivity of the composites
prepared by embedding spherical Ag particles (a mean radius $
\approx 100$  \AA\ ) into a KCl matrix.
The particles were made by
evaporating Ag in the presence  of argon and oxygen gases  so as
to form a thin (according to the authors, of approximately $10$
\AA\ , $\delta \simeq 0.10 $) oxide coating on them. This coating
prevented the particles from cold-welding together, but was thin
enough to allow metal-to-metal contact under high pressure. The
composites were prepared by mixing Ag particles and KCl powder and
then compressing the mixture into a solid pellet under high
pressure.
\begin{figure}[!b]
\centerline{\includegraphics[width=0.55\textwidth]{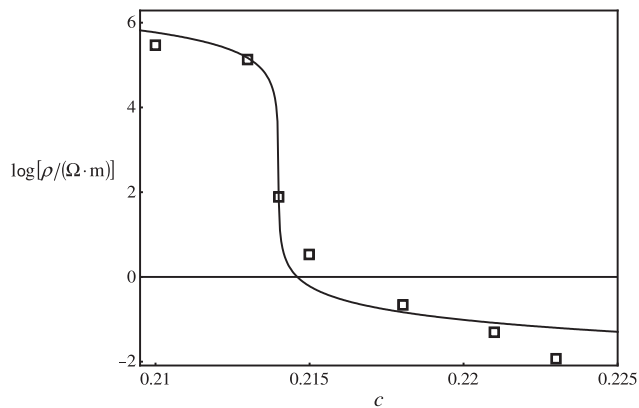}\hspace{3mm}}
\caption{Effective resistivity  data~\cite{bib:Che86} for  Ag--KCl
composite samples (squares) and the fit to them with formula
(\ref{cond1}) at $\sigma_1 = 6.3 \cdot 10^7$~{S/m},
$x_0=5\cdot 10^{-16}$, $\delta = 0.162$ ($c_{\rm c} = 0.214$), and
$x_2 = 4 \cdot 10^{-6}$.} \label{fig:Permit3e}
\end{figure}

It is seen from figure~\ref{fig:Permit3d} that formula (\ref{perm2}) not only
reproduces data~\cite{bib:Gra81} over the entire range of Ag
concentrations investigated, but also gives an estimate of $\delta
\simeq 0.14 \div 0.19$, sufficiently close to the expected one.

The conductivity (resistivity $\rho$) data for a few Ag--KCl
composite samples, prepared in the above way, are given in
\cite{bib:Che86}. They pertain only to a very narrow vicinity of
the percolation threshold, where $\rho$ drops 7 orders of
magnitude with a 1\% Ag volume concentration increment; the
parameters of the KCl matrix are not specified. As figure~\ref{fig:Permit3e} reveals, formula (\ref{cond1}) can reproduce
data~\cite{bib:Che86} sufficiently well. Better fits can be
produced by introducing $c$-dependences for some of the parameters
of the model. These facts may indicate that, in addition to
experimental errors, various other factors and phenomena
(inaccuracy of the function $\phi =\phi(c,\delta)$, particles'
size distribution, silver dissolution and local dielectric
breakdown in the KCl matrix, polarization effects, etc.) come into
play as $c_{\rm c}$ is approached. The analysis of them goes far
beyond the scope of this paper.

\subsection{``Double'' percolation}

\begin{figure}[h]
\centerline{\includegraphics[width=0.5\textwidth]{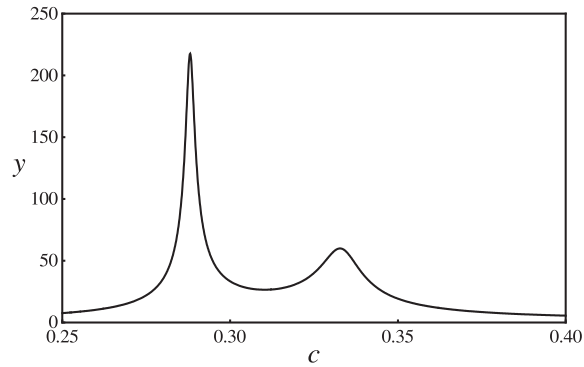}}
\caption{Effective permittivity as ``double'' percolation occurs;
 $x_0=1\cdot 10^{-8}$,
$x_2=5\cdot 10^{-4}$, $y_1= 1.5$, $y_2= 1$, $\delta =0.05$.}
\label{fig:Doub}
\end{figure}

For intermediate values of $x_2$ ($x_0\ll x_2 \ll x_1 $), a
``double'' percolation can be noticeable, that is, a new increase
in $x$ after some levelling off (figure~\ref{fig:Perc}); it is
accompanied by the appearance of a new peak in the concentration
dependence of the permittivity (figure~\ref{fig:Doub}). The
physical cause of this phenomenon is clear~--- in a concentrated system, the hard cores of particles with
penetrable shells begin to contact intensively to form percolation
cluster and add to the effective conductivity.

Evidently, the threshold concentration $c'_{\rm c}$ for ``double''
percolation is close to a value of 1/3. In the region $|c-
\frac{1}{3}| \ll x_2 \ll 1$, the $c$-dependence of the effective
conductivity (\ref{thr0}) is represented by the square-root law
\begin{equation} \label{thr6} x = {\frac{1}{2} \left(3 x_2
\right)^{{1}/{2}} \left[\phi(c,\delta)-\frac{1}{3}\right]
}^{{1}/{2}} + O(x_2) = \frac{1}{2} \left[3 x_2 \phi'(c_{\rm
c},\delta)  \right]^{{1}/{2}} \left(c - c_{\rm c}\right)^{1/2} +
O(x_2),
\end{equation}
$\phi'$ being the derivative of $\phi$  with respect to $c$. For
concentrations satisfying the condition $c - \frac{1}{3} \gg x_2$,
it becomes linear,
\begin{equation} \label{thr7} x = \frac{3}{2}
\left(c - \frac{1}{3}\right) + O(x_2),
\end{equation}
with a considerably greater amplitude  as compared to those  in
formulas (\ref{thr4}) and (\ref{thr6}).

As for now, we are unaware of experimental observations of the
effect described. Usually, ``double'' percolation is associated
with a non-monotonous behavior of the conductivity in composites
made by embedding the conducting particles into  a two-component
matrix (see, for example,~\cite{bib:Al08,bib:Kon06}).

It can also be shown that a decrease in $x$ can occur at  $c \sim
c'_{\rm c}$ in systems with $x_2\gg 1 $. Similar effects were
observed in two-phase composite solid electrolytes with a highly
conducting interphase layer~\cite{bib:Nan93}.

\section{Concluding remark}

The model proposed is interesting in the sense that  it is based
on rather clear assumptions, incorporates many-particle effects in
a consistent way, and can be further refined so as to apply to a
more complicated systems, including fluctuation phenomena, etc. For
practice, it can serve as a rather flexible theoretical basis for
analysis of dielectric and conductive properties of such complex
systems as various dispersions, colloids, and so-called
nanofluids, or for the development of new composite materials.

\section*{Acknowledgement}

We are grateful to an anonymous Referee for stimulating remarks.

\ukrainianpart
\title{Провідність та діелектрична проникність дисперсних систем із
вільно-проникним міжфазним шаром між частинками й середовищем}
\author{М.Я.~Сушко, А.К.~Семенов}
\address{Одеський національний університет імені І. І. Мечникова, вул.
Дворянська, 2, 65026 Одеса, Україна}

\makeukrtitle

\begin{abstract}
\tolerance=3000%
Запропоновано модель для вивчення ефективних квазістатичних
провідності та діелектричної про\-ник\-нос\-ті дисперсних систем із
вільно-проникним міжфазним шаром між частинками та середовищем,
яка ефективно враховує багаточастинкові поляризаційні та
кореляційні ефекти. Структура компонентів сис\-те\-ми, включаючи
міжфазний шар, враховується шляхом моделювання профілів їх
низькочастотної комп\-лекс\-ної діелектричної проникності. Модель,
зокрема, описує перколяційно-подібну поведінку ефективної
провідності, що супроводжується суттєвим зростанням дійсної
частини ефективної комплексної про\-ник\-нос\-ті системи. Положення
порогу перколяції визначається головним чином товщиною міжфазного
шару. Передбачено ефект  ``подвійної'' перколяції. Результати
порівняно з експериментом.

\keywords структурована частинка, дисперсна система, діелектрична
проникність, провідність, перколяція
\end{abstract}

\end{document}